# Naked Reissner-Nordström singularity and quasiclassical bound states

Valentin D. Gladush, Dmitry A. Kulikov

*Theoretical Physics Department, Dniepropetrovsk National University,*
*Gagarin Ave. 72, Dniepropetrovsk 49010. Ukraine*

**Abstract.** *The bound-state problem for an uncharged massive scalar particle in the field of a naked Reissner-Nordström singularity is approached by means of the quasiclassical Bohr-Sommerfeld quantization. An approximate analytical expression for the energy levels of the system is derived. It is shown that in order to obtain non-vanishing binding energies the masses of both the central object and the particle should be comparable with the Planck mass.*

**Introduction.** Quantum description plays an essential role in black hole physics and reveals some features that are unexpected from the viewpoint of classical theory. In particular, it is well established that the quantum spectrum of the Schwarzschild black hole does not contain stable bound states [3]. Instead, there exist resonance states decaying because of tunneling. The same occurs for the sub-extremal Reissner-Nordström (RN) blach hole [3] which carries not only the mass $M$ but also the charge $Q$ such that $Q^2 < \kappa M^2$ where $\kappa$ is the gravitational constant.

The main goal of the present work is to explore whether a quantum bound-state spectrum may arise in the super-extremal RN geometry with $Q^2 > \kappa M^2$. This geometry corresponds not to a black hole but to a naked singularity whose presence, at the classical level, results in the repulsive potential barrier near the origin [4; 9] that prevents the absorbtion of uncharged test particles in the quantum picture [1]. Thus, in view of the attractive nature of the Kepler potential tail, one may hope to obtain stable bound states.

Actually, the situation is more involved as the very existence of the naked RN singularities is questionable. It has been argued that they cannot be formed in the gravitational collapse [8] whereas the 'God given' RN singularities are neutralized by spontaneous pair creation provided that the charges are sufficiently high [2]. Moreover, the Klein-Gordon equation for a test particle or wave on the naked RN background suffers from ambiguity in choosing the boundary condition at origin [6].

In these circumstances it seems to be practical to approach the naked RN bound-state problem by means of the Bohr-Sommerfeld (BS) quantization. It does not require specifying the behaviour at origin and may even serve as an approximation for those systems in which the exterior naked RN geometry (where the particle moves) merges a different interior solution. Evaluating the BS integral, we calculate the discrete energy spectrum of the uncharged scalar test particle and estimate what parameter values are needed to obtain reasonable binding energies.

**Bohr-Sommerfeld quantization.** For writing down the quantization rule, we should derive the radial momentum for the test particle on the naked RN geometry. The metric of the naked RN manifold ($t \in R$, $r \in (0,\infty)$, $\Omega \in S^2$) is

$$ds = F(r)c^2 dt^2 - \frac{1}{F(r)} dr^2 - r^2 d\Omega^2, \quad F(r) = 1 - \frac{2\kappa M}{c^2 R} + \frac{\kappa Q^2}{c^4 R^2}, \quad d\Omega^2 = d\theta^2 + \sin\theta d\phi^2 \quad (1)$$

where the charge and the mass satisfy $Q^2 > \kappa M^2$. We choose $Q > 0$. The action for the uncharged scalar particle of mass $m$ is given by

$$S = -mc \int ds = \int \Lambda dt, \quad \Lambda = -mc\sqrt{F(r)c^2 - F^{-1}(r)\dot{r}^2 - r^2(\dot{\theta}^2 + \sin\theta\,\dot{\phi}^2)} \quad (2)$$

where $\dot{r} = dr/dt$ etc. Applying the Legendre transformation to $\Lambda$, we get the Hamiltonian

$$H = p_r \dot{r} + p_\theta \dot{\theta} + p_\phi \dot{\phi} - \Lambda = c\sqrt{F(r)\left[m^2 c^2 + F(r) p_r^2 + \frac{1}{R^2}\left(p_\theta^2 + \frac{p_\phi^2}{\sin^2 \theta}\right)\right]} \quad (3)$$

Since the quantities $H$ and $p_\theta^2 + p_\phi^2/\sin^2\theta$ are conserved, for our purposes we can replace them by their values: the total energy of particle $E$ and the angular momentum $L^2$, respectively. Then, for the radial momentum, we have

$$p_r^2 = \frac{1}{F(r)}\left(\frac{E^2}{F(r)c^2} - m^2 c^2 - \frac{L^2}{R^2}\right). \quad (4)$$

This is to be compared with the outcome of the massive Klein-Gordon equation

$$\left[\frac{1}{\sqrt{-g}} \frac{\partial}{\partial x^\alpha}\left(\sqrt{-g}\, g^{\alpha\beta} \frac{\partial}{\partial x^\beta}\right) + \frac{m^2 c^2}{\hbar^2}\right]\Psi = 0 \quad (5)$$

whose radial part, by virtue of substitution $\Psi = \exp(-iEt/\hbar) Y_l^m(\Omega) u(r)/r\sqrt{F(r)}$, reads

$$-\hbar^2 \frac{d^2 u}{dr^2} = \frac{1}{F(r)} \left( \frac{1}{F(r)} \left( \frac{E^2}{c^2} - \frac{\hbar^2 \kappa (Q^2 - \kappa M^2)}{c^4 R^4} \right) - m^2 c^2 - \frac{\hbar^2 l(l+1)}{R^2} \right) u. \tag{6}$$

We see that the only term in (6) that is absent in (4) is the right-hand-side term proportional to $\hbar^2\kappa$, which can be regarded as a higher-order quantum correction.

In order to obtain the discrete energy spectrum, we employ the BS quantization rule [7]

$$\oint p_r dr = 2\pi\hbar(n_r + 1/2) \tag{7}$$

where $n_r = 0,1,2,\ldots$ is the radial quantum number and, as it is usual in quasiclassics, we assume that the angular momentum inside of $p_r^2$ is defined with the Langer correction [7], that is $L = \hbar(l+1/2)$ ($l$=0,1,2,…).

It is convenient to introduce dimensionless quantities

$$\varepsilon = \frac{E}{mc^2}, \quad \sigma = \frac{\sqrt{\kappa} m}{cL} Q, \quad \mu = \frac{\sqrt{\kappa} m}{cL} M, \quad z = \frac{mc}{L} r. \tag{8}$$

In their terms the quantization rule becomes

$$\oint \frac{dz}{F(z)} \sqrt{\varepsilon^2 - W^2(z)} = \frac{2\pi(n_r + 1/2)}{l + 1/2} \tag{9}$$

with $F(z) = 1 - 2\mu/z + \sigma^2/z^2$ and $W^2(z) = (1 + 1/z^2)F(z)$ being the effective potential.

For integral (9) to be well-defined it is necessary to have two turning points, so that the integration contour may enclose them. This is fulfilled if the effective potential is single-well. Then the equation $dW/dz = 0$, which is cubic with respect to $z$, must have only one real root and its discriminant $D$ must be negative. Calculating

$$D = \frac{1}{\mu^4} \left( \mu^4 - \frac{1}{12} \left( 1 + 14\sigma^2 + \sigma^4 \right) \mu^2 + \frac{2\sigma^2}{27} (1+\sigma^2)^3 \right) \tag{10}$$

we find that $D < 0$ and $W(z)$ is indeed single-well for sufficiently light central objects with masses obeying

$$\mu^2 < \frac{2}{5}\sigma^2 \quad \text{that is} \quad \kappa M^2 < \frac{2}{5} Q^2. \tag{11}$$

It can be shown that when we approach the extremal case, so that $\kappa M^2$ increases and goes to $Q^2$, the effective potential eventually becomes double-well. In what follows, we treat only the case in which condition (11) is fulfilled.

**Calculation of energy spectrum.** We intend to obtain the energy levels of the particle in the naked RN geometry by evaluating the Bohr-Sommerfeld integral. Before doing this, let us estimate the quantity $\mu$ defined in (8). Rewriting

$$\mu = \frac{1}{l+1/2} \frac{m}{m_{Pl}} \frac{M}{m_{Pl}} \tag{12}$$

where $m_{Pl} = \sqrt{\hbar c/\kappa} = 2.18 \cdot 10^{-5}$ g is the Planck mass, we observe that, if the central object mass $M$ and the particle mass $m$ are of the elementary-particle scale, then $\mu$ is exceedingly small and thus the attractive $1/z$-term in the effective potential vanishes. We therefore assume that $M$ is of the same order as $m_{Pl}$ whereas $m$ is of a few orders less so as to satisfy $m \ll M$. Then it holds $\mu \ll 1$, but $\mu$ cannot be neglected.

For calculating the integral (9) in an approximate way, we rescale to the new coordinate $y$ and the new energy parameter $\lambda$, by substituting $z = y/\mu$ and $1 - \varepsilon^2 = \lambda \mu^2$. Retaining only those terms that are independent of $\mu$, we arrive at the integral

$$\oint \frac{dy}{y} \sqrt{-\lambda y^2 + 2y - 1 - \sigma^2} = \frac{2\pi(n_r + 1/2)}{l + 1/2}, \tag{13}$$

which is readily evaluated analytically. In fact, its functional dependence on $y$ coincides with that of the BS integral in the non-relativistic Coulomb problem. Since for the Coulomb problem the BS quantization reproduces the exact expression for eigenenergies, it seems to be justified in the naked RN bound-state problem, too.

As a result, we get the approximate formula for energy levels of the particle on the naked RN background

$$\frac{E}{mc^2} = \left[ 1 - \left( \frac{\kappa m M}{\hbar c} \right)^2 \Bigg/ \left( n_r + \frac{1}{2} + \sqrt{\left( l + \frac{1}{2} \right)^2 + \frac{\kappa m^2}{\hbar c} \cdot \frac{Q^2}{\hbar c}} \right)^2 \right]^{1/2}. \tag{14}$$

For comparison, we write down the well-known expression for eigenenergies of the Klein-Gordon particle with charge $-q$ and mass $m$ placed in the Coulomb field, $Q/r$, in the flat spacetime [5]

$$\frac{E}{mc^2} = \left[1 + \left(\frac{qQ}{\hbar c}\right)^2 \Big/ \left(n_r + \frac{1}{2} + \sqrt{\left(l+\frac{1}{2}\right)^2 - \left(\frac{qQ}{\hbar c}\right)^2}\right)^2\right]^{-1/2}. \tag{15}$$

The similarity of these two formulas becomes transparent in the non-relativistic limit in which they both yield

$$E_{nonrel} = E - mc^2, \quad \frac{E_{nonrel}}{mc^2} = -\frac{\beta^2}{2(n_r + l + 1)^2} \tag{16}$$

with β standing for $\kappa mM/(\hbar c)$ and $qQ/(\hbar c)$ in the case of (14) and (15), respectively. Rewriting $\kappa mM/(\hbar c)$ as $(m/m_{Pl})(M/m_{Pl})$, we see once again that in the RN case the masses should be comparable with the Planck mass in order to get non-vanishing binding energies. Let us recall, however, that our consideration refers to the central object mass restricted by (11) so as to have the single-well effective potential. For the near-extremal naked RN singularity the potential is double-well and a further analysis is needed to establish whether one may bind lighter particles.

It should be noticed that the Coulomb formula (15) implies that the charge $Q$ may not be unrestrictedly large and there must exist its critical value. Physically, it is caused by spontaneous pair production in strong electromagnetic field. Although in itself the RN formula (14) does not restrict the charge of the central object, this is merely due to the fact that the test particle is uncharged. For charged particles, it has been shown [2] that in the naked RN geometry with $Q \gg 137e$ where $e$ is the elementary charge the pair production occurs as well.

**Numerical results.** For estimating the accuracy of our approach, we compare the values of energies $E_{BS}$ calculated using the approximate formula (14) with the values $E_{num}$ obtained by direct numerical integration of the Klein-Gordon equation (6). Since the latter requires the mixed boundary condition at origin $du(0)/dr + a\,u(0) = 0$ with an arbitrary real value of the parameter $a$ [6], we consider two extreme cases: $a = 0$ and $a = \infty$.

In Table 1 the computed eigenenergies of the first s-states ($l = 0$) are given. The calculation has been performed with $Q = 30e$ and $M = 1.5 m_{Pl}$ (so that $\kappa M^2/Q^2 = 0.34 < 2/5$ in accordance with (11)) for the two values of the particle mass: $m = M/25$ and $m = M/50$. From this table it is evident that the eigenenergies are insensitive to the choice of the boundary condition at origin, especially when the particle is lighter. This is because the allowed region of the particle radial coordinate in the classical theory is located far from the origin. As seen from Table 1, the BS integral over the allowed region results in the accurate values of energies. The agreement is better for excited states, as expected in quasiclassics. Remarkably, all these values are close to the rest energy, that is the system is in fact non-relativistic.

*Table 1. Energies of the uncharged scalar particle in the naked RN geometry*

| $n_r$ | $m = M/25$ | | | $m = M/50$ | | |
|---|---|---|---|---|---|---|
| | $E_{BS}$ | $E_{num}$ | | $E_{BS}$ | $E_{num}$ | |
| | | $a = 0$ | $a = \infty$ | | $a = 0$ | $a = \infty$ |
| 0 | 0.996123 | 0.995976 | 0.995955 | 0.998999 | 0.998988 | 0.998986 |
| 1 | 0.999010 | 0.998983 | 0.998979 | 0.999748 | 0.999747 | 0.999746 |
| 2 | 0.999557 | 0.999548 | 0.999547 | 0.999888 | 0.999887 | 0.999887 |

**Conclusion.** We have treated the quantum bound-state problem for the uncharged massive scalar particle in the field of the naked RN singularity by using the quasiclassical BS approach. This approach enabled us to avoid the ambiguity in choosing the boundary condition at origin and to derive the approximate formula for discrete energy levels of the system. The calculated binding energies turn out to be non-vanishing if the masses of both the central object and the particle are comparable with the Planck mass. A possibility to relax the restriction on the particle mass may consist in studying the case of the near-extremal naked RN singularity.